# Revealing interstitial energetics in Ti-23Nb-0.7Ta-2Zr gum metal base alloy via universal machine learning interatomic potentials

Miroslav Lebeda[a,b,c], Jan Drahokoupil[a], Veronika Mazáčová[a, c], Petr Vlčák[a]

[a] *Faculty of Mechanical Engineering, Czech Technical University in Prague, Technická 4, 16607 Prague 6, Czech Republic*

[b] *Faculty of Nuclear Sciences and Physical Engineering, Czech Technical University in Prague, Trojanova 339/13, 12000 Prague 2, Czech Republic*

[c] *FZU – Institute of Physics of the Czech Academy of Sciences, Na Slovance 2, 18200 Prague 8, Czech Republic*

Corresponding author: Miroslav Lebeda, lebedmi2@cvut.cz

## Abstract

Understanding the behavior of light interstitial elements in multicomponent alloys remains challenging due to the complexity of local chemical environments and the high computational cost of first-principles calculations. Here we demonstrate that three universal machine-learning interatomic potentials (uMLIPs) – MACE-MATPES-PBE-0, Orb-v3, and SevenNet-0 can efficiently map the energetics of C, N, O, and H interstitials in a Ti-23Nb-0.7Ta-2Zr (at.%) gum metal base alloy while being several orders of magnitude faster than density functional theory (DFT). All uMLIPs predict broad energy distributions (~1-3 eV) across the four interstitial elements, reflecting their strong sensitivity to local lattice chemistry. Despite alloy disorder, MACE-MATPES-PBE-0 and Orb-v3 reproduce the expected site preferences of the bcc structure: C, N, and O relax into octahedral sites, whereas H stabilizes in tetrahedral positions. In contrast, SevenNet-0 predicts H to be most stable in octahedral coordination, indicating a limitation of this model. Correlation analysis reveals two dominant chemical trends: Ti-rich environments strongly stabilize interstitials, whereas close proximity to Nb is destabilizing. Zr and Ta show no statistically significant influence, likely due to their low concentrations. Benchmarking representative O interstitial configurations against DFT confirms that the uMLIPs reasonably reproduce the energetic ordering of chemically distinct environments. Additional DFT validation confirms that tetrahedral configurations are energetically more favorable than octahedral sites for H interstitials, further illustrating the SevenNet-0 limitation. Overall, these results demonstrate that uMLIPs enable computationally efficient, statistically broad characterization of defect energetics in Ti-23Nb-0.7Ta-2Zr gum metal base alloy and provide insight into how local chemical environments govern interstitial stability.

## 1. Introduction

Middle- and high-entropy alloys (M/HEAs) and other compositionally complex alloys typically exhibit exceptional combinations of strength, ductility, and thermal stability, largely due to the wide variations of local chemical environments present in their disordered atomic structures[1–4]. In bcc Ti-Nb-Zr based systems, light interstitials such as N, C, O, and H strongly influence phase stability and mechanical properties, even at low concentrations[5–9]. The interstitial behavior is highly affected by the chemistry and geometry of the surrounding host atoms, making quantitative understanding of interstitial energetics important for interpretation and modifications of alloy performance.

A particularly relevant case where interstitials play a central functional role is Ti-23Nb-0.7Ta-2Zr (at.%) gum metal base alloy. Its characteristic combination of high strength (> 1 000 MPa), low elastic modulus (~ 50 GPa), and large recoverable strains (up to 2.5 %) is obtained only within a narrow window of O content, where O provides substantial interstitial strengthening[10–12]. Other light interstitials may play similarly important roles, yet their energetics in the gum metal base alloy matrix remain largely unexplored[13].

Although the behavior of light interstitials in complex alloys has been studied with first-principles methods[14–16], such analyses are typically statistically limited: realistic modeling requires reasonably large special quasirandom structures (SQSs)[17,18] to properly represent chemical disorder, and the number of available interstitial configurations grows rapidly with supercell size, making the DFT sampling computationally too expensive. As a result, a broader statistical analysis of interstitial energetics in the gum metal base alloy remains an open question.

Machine-learning interatomic potentials (MLIPs) have emerged as a promising approach to overcome DFT limitations. Yet, most existing MLIPs are system-specific[19], requiring dedicated training for each alloy composition or interstitial elements, which restricts their transferability. Recently, universal MLIPs (uMLIPs, sometimes called foundation models), such as CHGNet[20], M3GNet[21], MACE[22,23], Orb[24], MatterSim[25], SevenNet[26], Nequix[27], or PET-MAD[28] trained on various DFT datasets[29–33] have demonstrated near-DFT accuracy across broad chemical elements. Compared with DFT, uMLIPs offer orders of magnitude faster simulation times, enabling calculations on large supercells and far more atomic configurations while also reducing dependence on high-performance computing (HPC). Consequently, uMLIPs open new opportunities for modeling complex alloys at scales previously unattainable by first-principles methods[34].

In this work, we employ three selected uMLIPs (MACE-MATPES-PBE-0, Orb-v3, and SevenNet-0) to investigate the energetics of C, N, O, and H interstitials in the Ti-23Nb-0.7Ta-2Zr gum metal base alloy. Using three independent 250-atom SQSs to represent the alloy's substitutional disorder, we systematically evaluate all octahedral and tetrahedral interstitial sites (6750 in total) for all four interstitial elements. To ensure reproducibility and broaden accessibility, we developed uMLIP-Interactive (github.com/bracerino/uMLIP-Interactive), an open-source graphical user interface (GUI) designed to make simulations with uMLIP more user-friendly. The platform integrates a wide range of pretrained models, including MACE, CHGNet, SevenNet, Orb, MatterSim, Nequix, and PET-MAD, and currently enables to perform single-point energy calculations, structural relaxations, elastic-constant evaluations, phonon calculations, and basic molecular dynamics (MD) within an intuitive interface. The tool also automatically generates ready-to-run all-in-one Python scripts that reproduce the chosen simulation setup, which allows easily shareable calculation scripts.

Our results reveal clear chemical trends across all sampled interstitial configurations in Ti-23Nb-0.7Ta-2Zr gum metal base. Local environments enriched in Ti consistently stabilize N, C, O, and H interstitials, whereas proximity to Nb strongly destabilizes them. Zr and Ta exhibit statistically insignificant influence, likely due to their low concentrations in the gum metal base alloy. Despite extensive substitutional disorder, the typical geometric preferences of the bcc lattice persist for MACE-MATPES-PBE-0 and Orb-v3, with N, C, O relaxing into octahedral sites and H remaining tetrahedral. However, SevenNet-0 predicts an exception for H, stabilizing it into octahedral coordination. This illustrates SevenNet-0 possible limitation for the H description. Comparing randomly selected configurations with DFT data supports the general trends observed by uMLIPs. Overall, our findings show that uMLIPs can reliably capture how local chemistry influences interstitial energetics while being several orders of magnitude faster than DFT, enabling statistically converged analyses that would not be typically possible with first-principles methods alone.

## 2. Methods

### 2.1 SQS generation

The SQS for Ti-23Nb-0.7Ta-2Zr gum metal base alloy were generated using the ATAT mcsqs module[18,35]. The initial structure for the bcc phase was based on β-Ti (Im-3m, $a$ = 3.282 Å[36]). A 250-atom supercell of dimensions 5 × 5 × 5 was constructed, employing pair clusters up to 2.0 Å and triplet clusters up to 1.5 Å (normalized relative to the maximum lattice parameter). For statistics, three best independent SQS configurations searched for 24 hours were selected for subsequent calculations. This supercell size allows to approximate the gum metal base alloy composition of 74.3Ti-23Nb-0.7Ta-2Zr (at.%) as 74Ti-23.2Nb-0.8Ta-2Zr (at.%).

For each SQS structure, all tetrahedral (1500 sites) and octahedral (750 sites) interstitial positions were systematically populated with a single interstitial atom (0.4 at. %). Only one interstitial was introduced per supercell to avoid interstitial-interstitial interactions, so that the energies reflect intrinsic site stability governed solely by the local host-lattice chemical environment. Because three best SQS were used to improve statistics of different chemical environments around the interstitial elements, there were in total 6750 interstitial sites characterized. The entire SQS workflow generation was automated using an all-in-one bash script from SimplySQS[37] interface (simplysqs.com). The bash script is publicly available on GitHub at github.com/bracerino/Ti-23Nb-0.7Ta-2Zr-Interstitials.

### 2.2 DFT settings

First-principles calculations were performed using the Vienna Ab Initio Simulation Package (VASP)[38]. A plane-wave basis set with an energy cutoff of 620 eV was employed. Brillouin-zone sampling was restricted to the Γ-point. Exchange-correlation effects were treated within the generalized gradient approximation (GGA) using the Perdew-Burke-Ernzerhof (PBE) functional[39]. Electronic self-consistency was reached with an energy convergence threshold of $10^{-4}$ eV. Structural optimizations were performed with the conjugate-gradient algorithm, allowing both the atomic positions and the cell to relax while constraining the cell angles. Geometry optimizations were converged with an energy tolerance of $10^{-3}$ eV.

### 2.3 uMLIPs settings

Structural optimizations and energy evaluations of the interstitial configurations were performed within the Atomic Simulation Environment (ASE) framework[40] using Broyden-Fletcher-Goldfarb-Shanno (BFGS) optimizer and the following three uMLIPs: MACE-MATPES-PBE-0[22], Orb-v3[24], and SevenNet-0[26]. These models were chosen as representative potentials from three independently developed and widely used uMLIP families (MACE, Orb, and SevenNet). The force convergence criterion of 0.005 eV/Å was employed. During the structural optimization, both the atomic positions and the cell (with fixed angles) were allowed to relax.

To facilitate and standardize our simulations, we developed and used *uMLIP-Interactive* (github.com/bracerino/uMLIP-Interactive), an open-source graphical user interface (GUI) that integrates basic calculations with uMLIPs (illustrative video available at: youtu.be/xh98fQqKXaI?si=QhXIZs1JKYNmTnVq). The application allows users to import atomic structures in standard formats and either run simulations directly within the interface or generate all-in-one Python scripts for standalone execution. The supported simulation types currently include single-point energy calculations, geometry optimizations, elastic-property evaluations, phonons, and basic MD simulations. The option to create a self-contained Python script that can be executed in separated directory ensures easily reproducible and shareable calculations. Beyond research applications, this interface also can serve as an educational tool, allowing students and new users to explore uMLIPs simulations interactively. We provide all uMLIP-Interactive scripts used for the structural optimizations with the three uMLIPs also in the following repository: github.com/bracerino/Ti-23Nb-0.7Ta-2Zr-Interstitials.

# 3. Results

## 3.1 Energetics of interstitial configurations

Evaluating all structurally-optimized 6750 interstitial configurations (750 octahedral and 1500 tetrahedral sites per each of three independent SQS, see also **Fig. 1** for the employed workflow) with three uMLIPs (MACE-MATPES-PBE-0, Orb-v3, SevenNet-0) in Ti-23Nb-0.7Ta-2Zr reveals a wide total energy range of ~1 - 3 eV across all four interstitial elements (C, N, O, H). Additionally, uMLIPs clearly predict preferences between octahedral and tetrahedral sites (**Fig. 2**):

- **N, C, and O preferentially occupy octahedral sites:** Tetrahedral initial positions typically relax into the nearest octahedral minimum with all three uMLIPs. The few tetrahedral configurations that persist are consistently higher in energy than octahedral sites.
- **H preferentially occupies tetrahedral sites:** MACE-MATPES-PBE-0 and Orb-v3 mostly stabilize H in tetrahedral sites and reproduce the expected behavior for bcc alloys. In contrast, SevenNet-0 predicts that H is energetically favored in octahedral coordination and systematically relaxes tetrahedral starting positions into octahedral minima. This illustrates that SevenNet-0 is potentially insufficient as uMLIP for the description of H behavior in the studied gum metal base alloy. This discrepancy is further clarified by the DFT validation presented in Section 3.3, which confirms the energetic preference of tetrahedral configurations.

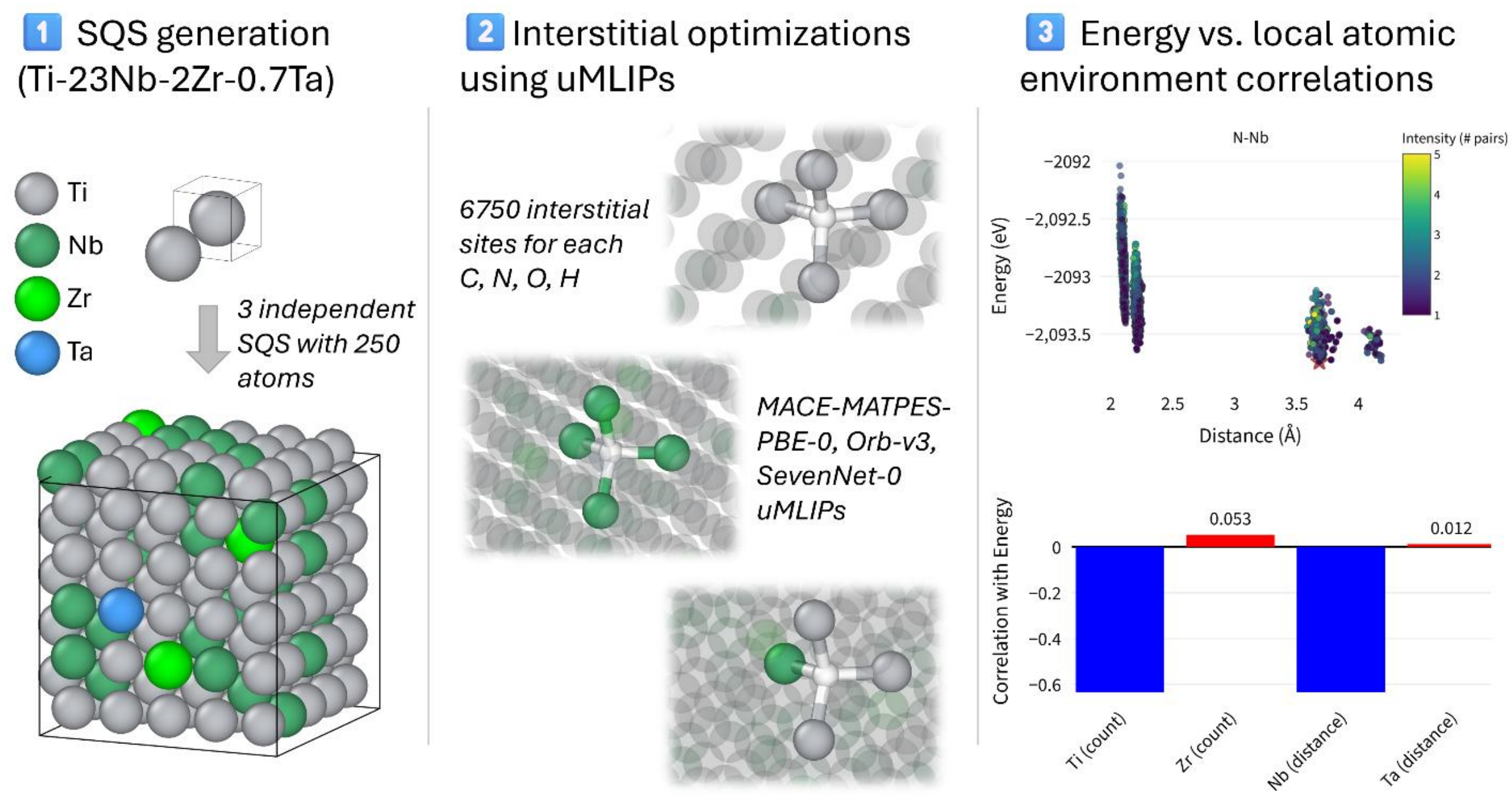

**Fig. 1:** Workflow used to statistically characterize the influence of local chemical environments in the Ti-23Nb-0.7Ta-2Zr gum metal base alloy on the energetics of C, N, O, and H interstitials using three uMLIPs.

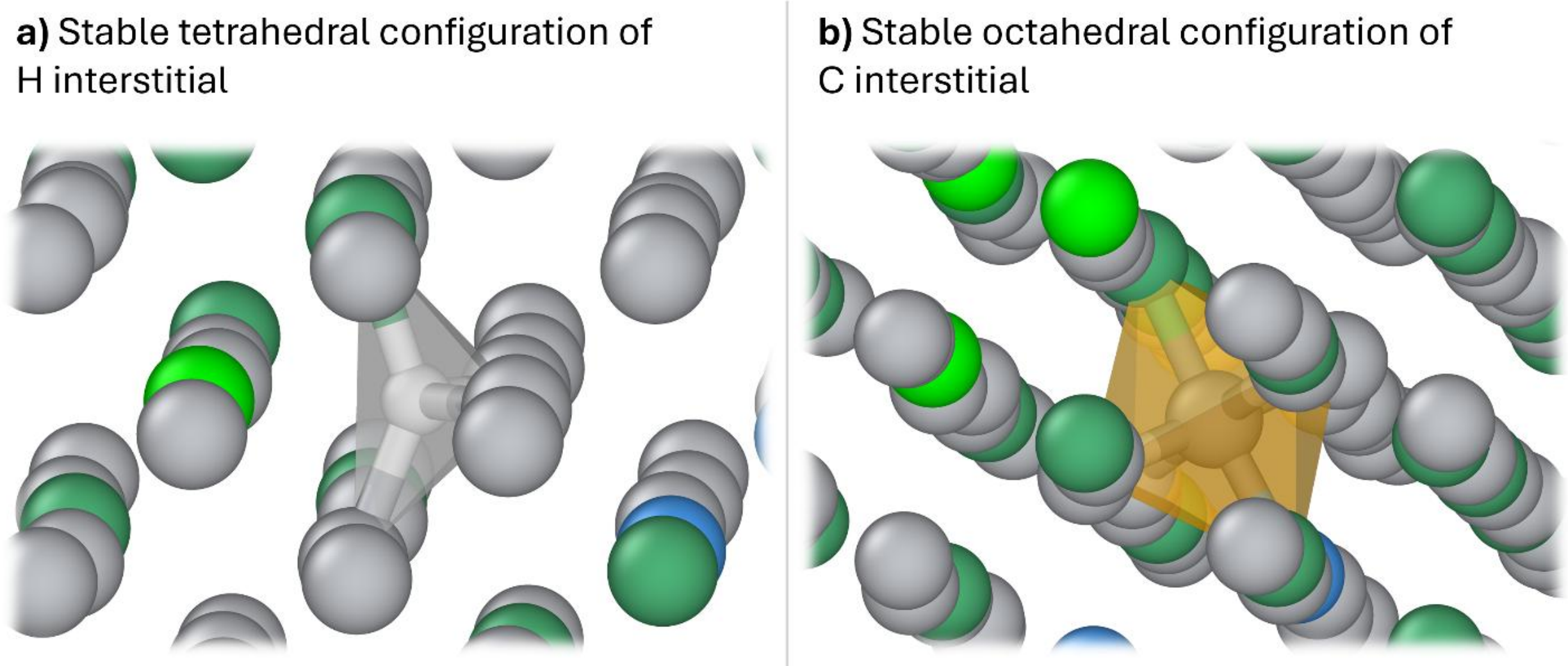

**Fig. 2:** Illustration of stable tetrahedral configuration for H interstitial (**a**) and stable octahedral configuration for N, C, O interstitials (**b**) after structure optimizations in Ti-23Nb-0.7Ta-2Zr with uMLIPs (MACE-MATPES-PBE-0, Orb-v3). SevenNet-0 uMLIP differs for H interstitial and relax it into octahedral sites.

## 3.2 Effect of local chemical environment on interstitial energetics

To identify the origin of the energy variations across all 6750 interstitial configurations, we quantified two descriptors for each interstitial: (**i**) the nearest-neighbor (NN) count of each host element (within 0.1 Å distance tolerance) and (**ii**) the minimum interstitial-host distance for each element. As illustrated for N interstitials in **Fig. 3**, qualitative trends are visible: configurations with a higher Ti NN count tend to exhibit lower energies (see heatmap values), whereas larger Nb-N distances likewise reduce energy. In contrast, descriptors associated with Ta and Zr show no apparent influence, likely due to their low concentrations in the Ti-23Nb-0.7Ta-2Zr alloy. Plots for all interstitials can be found in **Supplementary information**. To quantify these trends, we computed Pearson correlation coefficients ($r$) between each descriptor and the total energy. We interpret the coefficients as follows: negative correlations ($-1 < r < -0.1$) indicate that increasing the NN count of an element lowers the energy, or for the distance descriptor, that larger distances correspond to lower energies; positive correlations ($0.1 < r < 1$) reflect the opposite trend; and values within $-0.1 \leq r \leq 0.1$ are considered negligible.

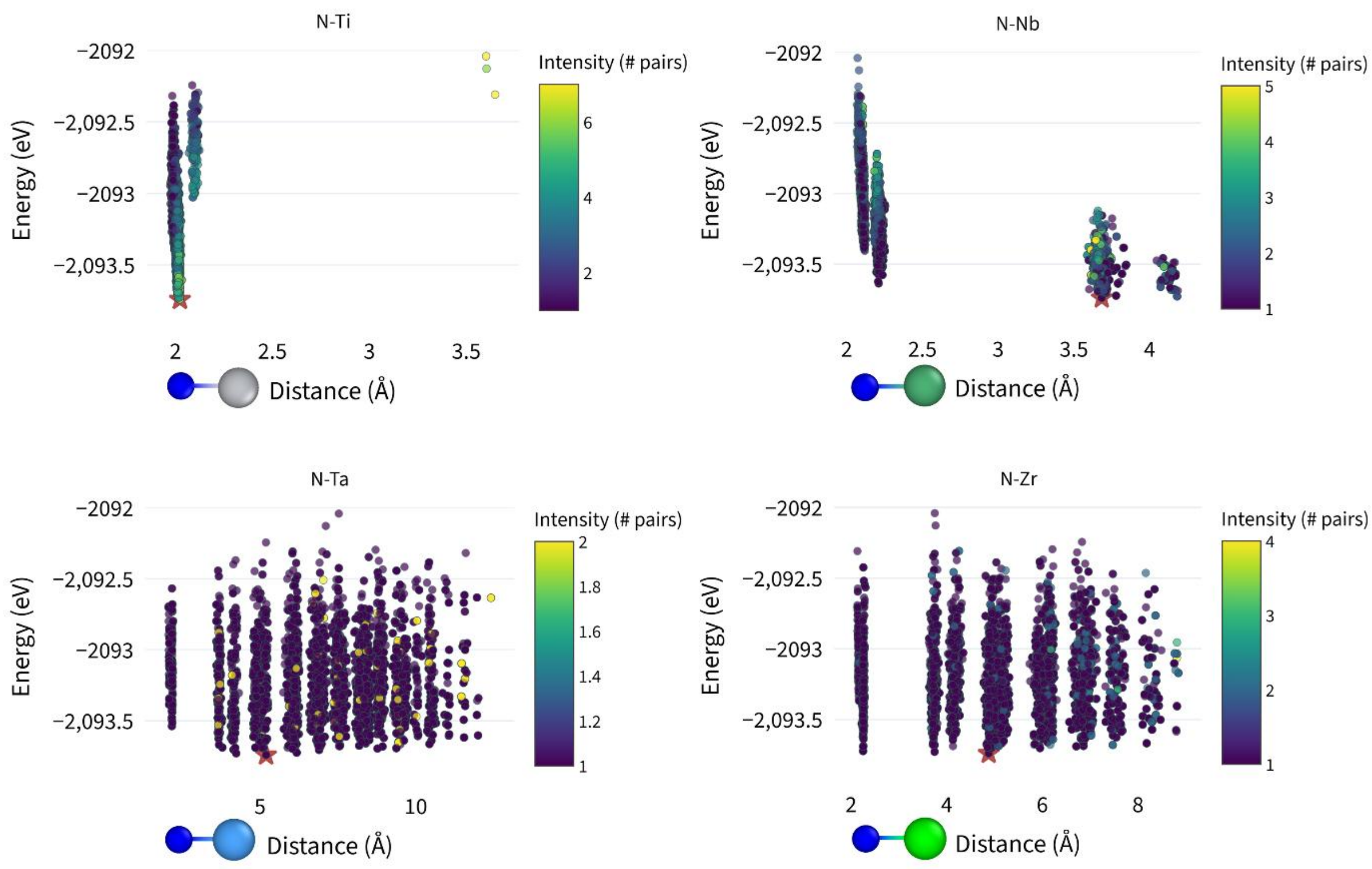

**Fig. 3**: MACE-MATPES-PBE-0 predicted energy of an N interstitial in Ti-23Nb-0.7Ta-2Zr as a function of nearest-neighbor (NN) distance, with the NN elemental count shown as a heatmap. The top left plot shows a clear correlation between increasing Ti NN count and decreasing interstitial energy, while the top right plot indicates that larger Nb NN distances also correspond to decrease in energy. In contrast, Ta and Zr (bottom two plots) show no visible correlation with the interstitial energetics. The star in the plots illustrates the overall lowest energy configuration.

Across all uMLIPs and all interstitial elements, the following consistent chemical trends appear:

1) **Ti-rich environments are stabilizing**

All three uMLIPs show that increasing the number of Ti NN lowers the interstitial energy, as indicated by negative correlations (**Fig. 4**). MACE-MATPES-PBE-0 exhibits the strongest Ti sensitivity, particularly for H, followed by N, C, and O. Orb-v3 predicts the largest Ti correlation for O and N, with weaker trends for the other elements. SevenNet-0 shows an ordering of C, N, H, but unlike the other models, it exhibits almost negligible correlation ($r \sim 0.1$) between Ti count and O energetics, indicating potential model specific limitation.

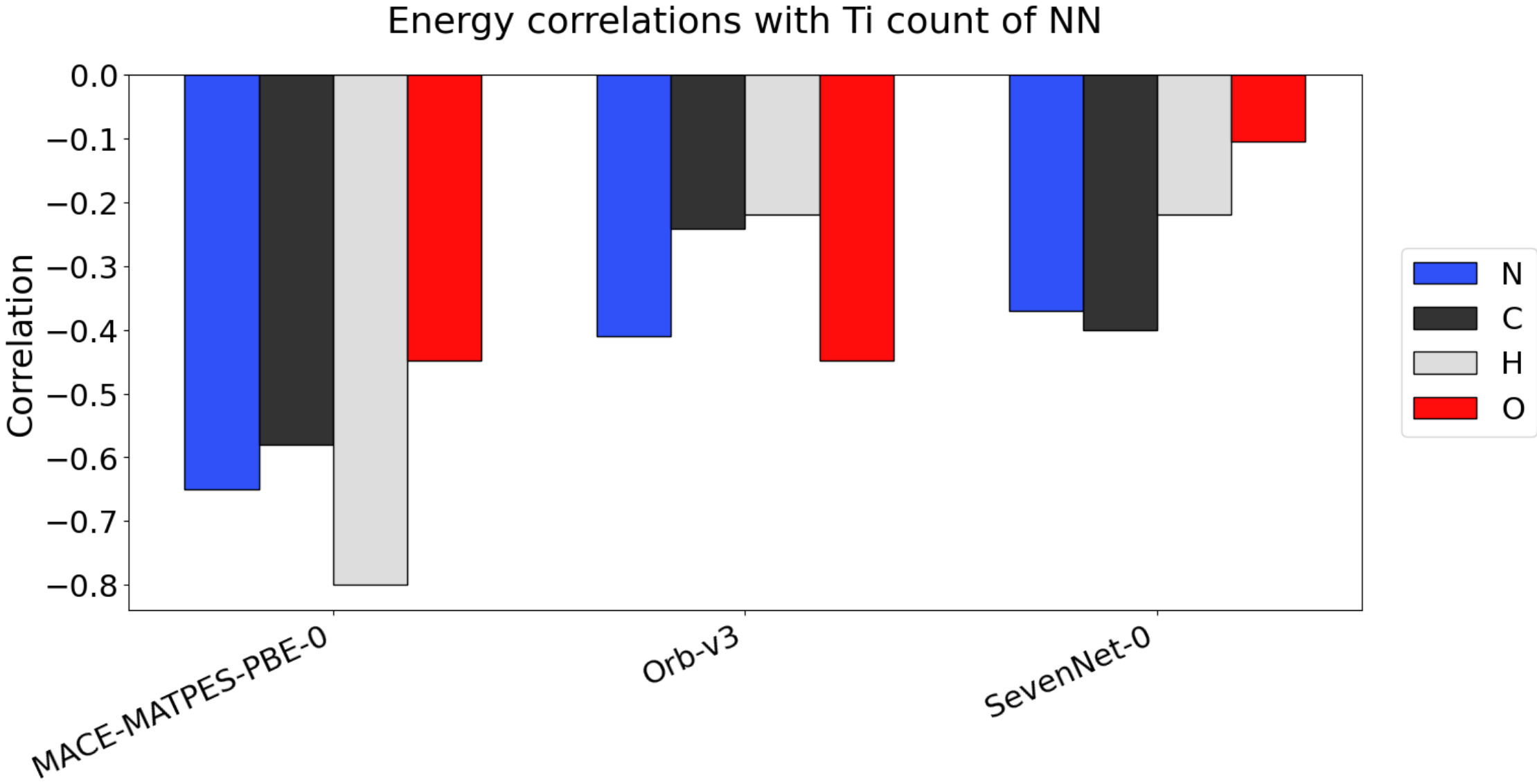

**Fig. 4:** Pearson correlation between the total energies of C, N, O, and H interstitial configurations in the Ti-23Nb-0.7Ta-2Zr gum metal base alloy and the number of Ti nearest neighbors around each interstitial site. Energies are predicted using the MACE-MATPES-PBE-0, Orb-v3, and SevenNet-0 uMLIPs. Higher Ti coordination correlates with lower interstitial energies across all models and interstitial elements.

### 2) Close Nb distances are destabilizing

Nb exerts the opposite effect: shorter Nb interstitial distances correlate with higher energies, while increasing this distance lowers the energy (**Fig. 5**). Thus, the destabilizing role of Nb manifests primarily through distance rather than NN count. MACE-MATPES-PBE-0 and SevenNet-0 display similar trend ordering, with O showing the strongest Nb sensitivity, followed by H, N, C. Orb-v3 likewise predicts the strongest correlation for O, but exhibits a different ordering for the remaining elements, with N more affected than C and H. Overall, all models agree that O is most strongly destabilized by close Nb proximity.

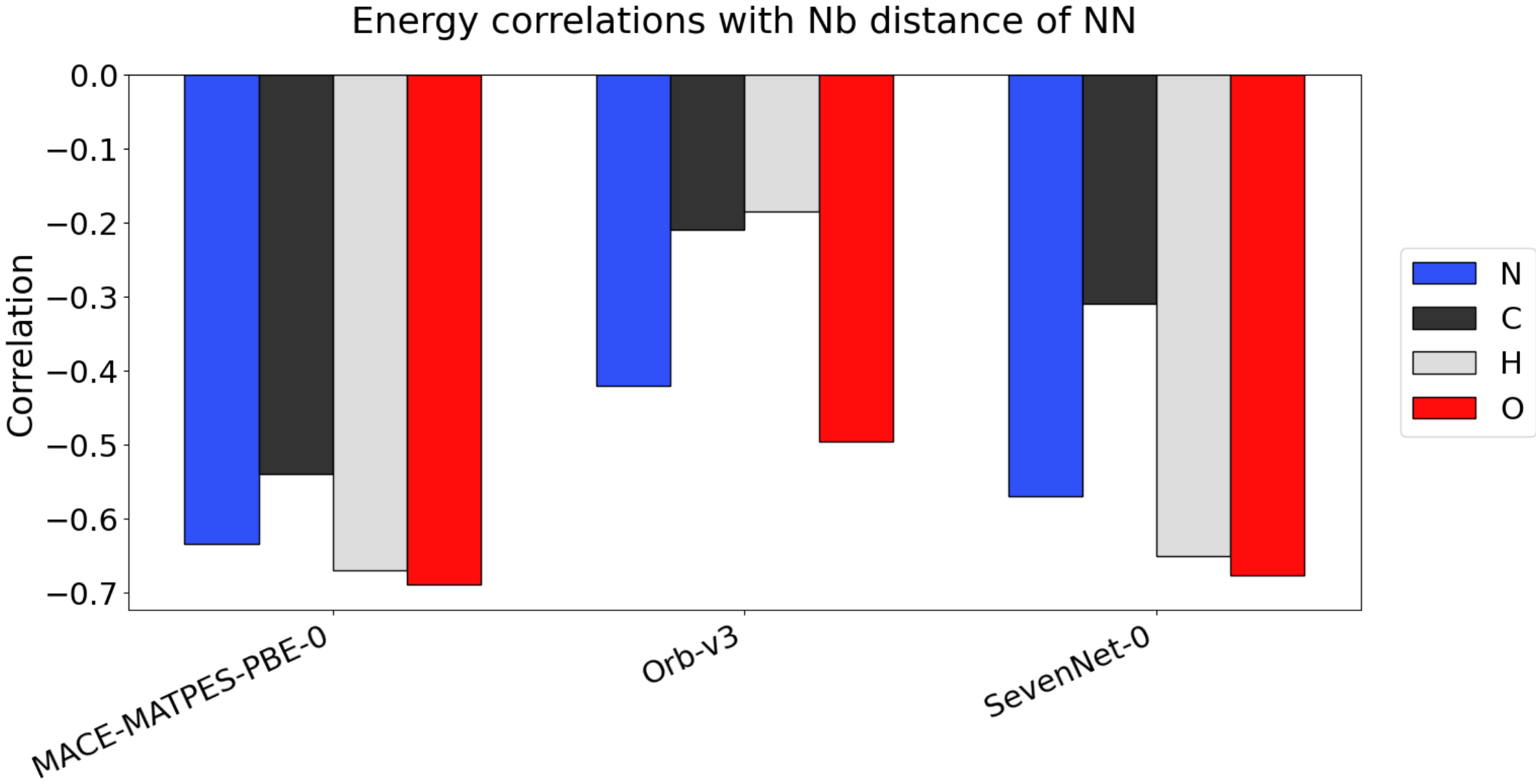

**Fig. 5:** Correlation between the total energies of C, N, O, and H interstitial configurations in the Ti-23Nb-0.7Ta-2Zr gum metal base alloy and the minimum Nb-interstitial distance. Energies are obtained using the MACE-MATPES-PBE-0, Orb-v3, and SevenNet-0 uMLIPs. All models show that shorter Nb-interstitial distances correlate with higher energies.

### 3) Zr and Ta exert negligible influence

Both Zr and Ta show no measurable correlation with interstitial energies.

Although the three uMLIPs exhibit consistent qualitative trends in how local chemical environments influence interstitial energetics, some quantitative differences remain. MACE-MATPES-PBE-0 shows overall the strongest correlations between the chemical descriptors and the energies, indicating high sensitivity to local bonding variations. Orb-v3, while reproducing the same chemical trends, yields the weakest quantitative correlations among the three models (with the exception for Ti count for O, where it is the lowest by SevenNet-0). Direct comparison of the energies predicted by each pair of uMLIPs using Pearson correlation coefficients confirms general differences between models: for C interstitials, MACE-MATPES-PBE-0 and SevenNet-0 correlate with $r = 0.415$, SevenNet-0 and Orb-v3 with $r = 0.407$, whereas MACE-MATPES-PBE-0 and Orb-v3 show the lowest agreement ($r = 0.246$).

### 3.3 DFT validation of uMLIP predictions

To validate the predictions of the uMLIPs, we randomly selected six O interstitial configurations spanning across the energetic range predicted by Orb-v3 and compared them with DFT and the other two uMLIPs. For each configuration, we evaluated (**i**) DFT single-point energies on the selected Orb-v3 relaxed geometries and (**ii**) fully DFT-relaxed energies. The static (single-point) DFT calculations reproduce the monotonic trend predicted by Orb-v3 for the three lowest-energy configurations (**Fig. 6a**). In contrast, the three higher-energy configurations exhibit pronounced deviations from linearity, and the highest energy pair (structures 5 and 6) shows a small inversion in ordering. This small inversion appears in a part of the energy landscape that is already far from the low-energy minima and therefore is expected to have limited impact on physically relevant behavior.

After full DFT geometry optimization, the overall energetic ordering remains consistent with the single-point comparison, although the absolute energies shift notably closer to the Orb-v3 predictions. When comparing the relative energies of the ranked structures across all three uMLIPs (**Fig. 6b**, shown as energy differences with respect to the lowest energy structure predicted by Orb-v3), generally both the lowest and highest energy configurations are reproduced by all models and by DFT. Some deviations appear in the intermediate energy structures. SevenNet-0 predicts structure 3 to lie slightly below structure 2, but correctly places structures 4 and 5 at higher energies and reproduces the small inversion between structures 5 and 6 observed in DFT. MACE-MATPES-PBE-0 similarly shifts the ordering, placing structures 3 and 4 slightly below structure 2, while still capturing the higher energies of structures 5 and 6 in agreement with Orb-v3.

Despite these differences, all uMLIPs identify similar low- and high-energy states in accordance with DFT. This demonstrates that the uMLIPs capture the energetic hierarchy of O interstitial configurations in Ti-23Nb-0.7Ta-2Zr reasonably well, even without system-specific training or fine-tuning. Moreover, the close agreement between uMLIP-relaxed and DFT-relaxed geometries indicates that uMLIPs provide high-quality starting structures for DFT optimizations, which can substantially accelerate DFT convergence.

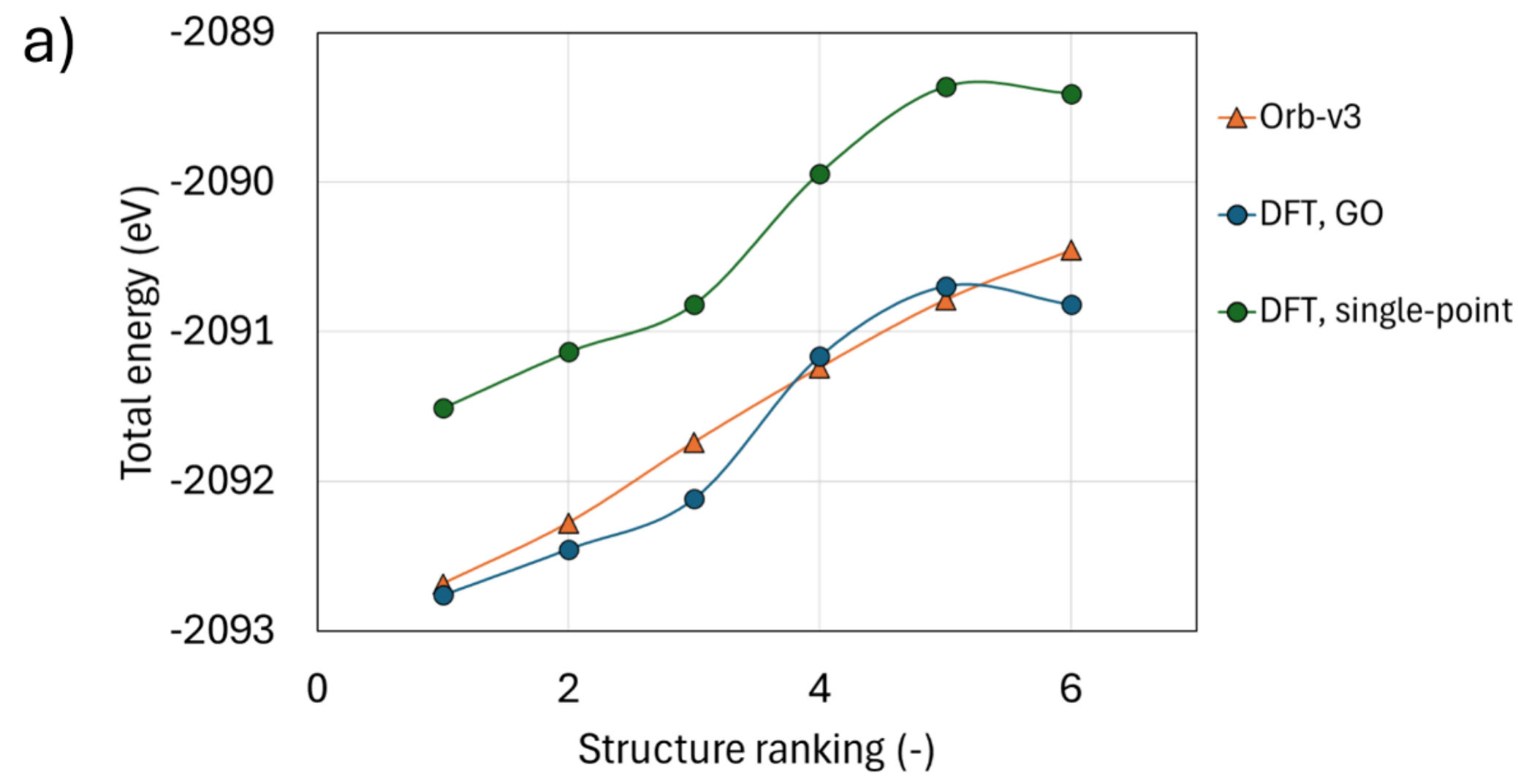

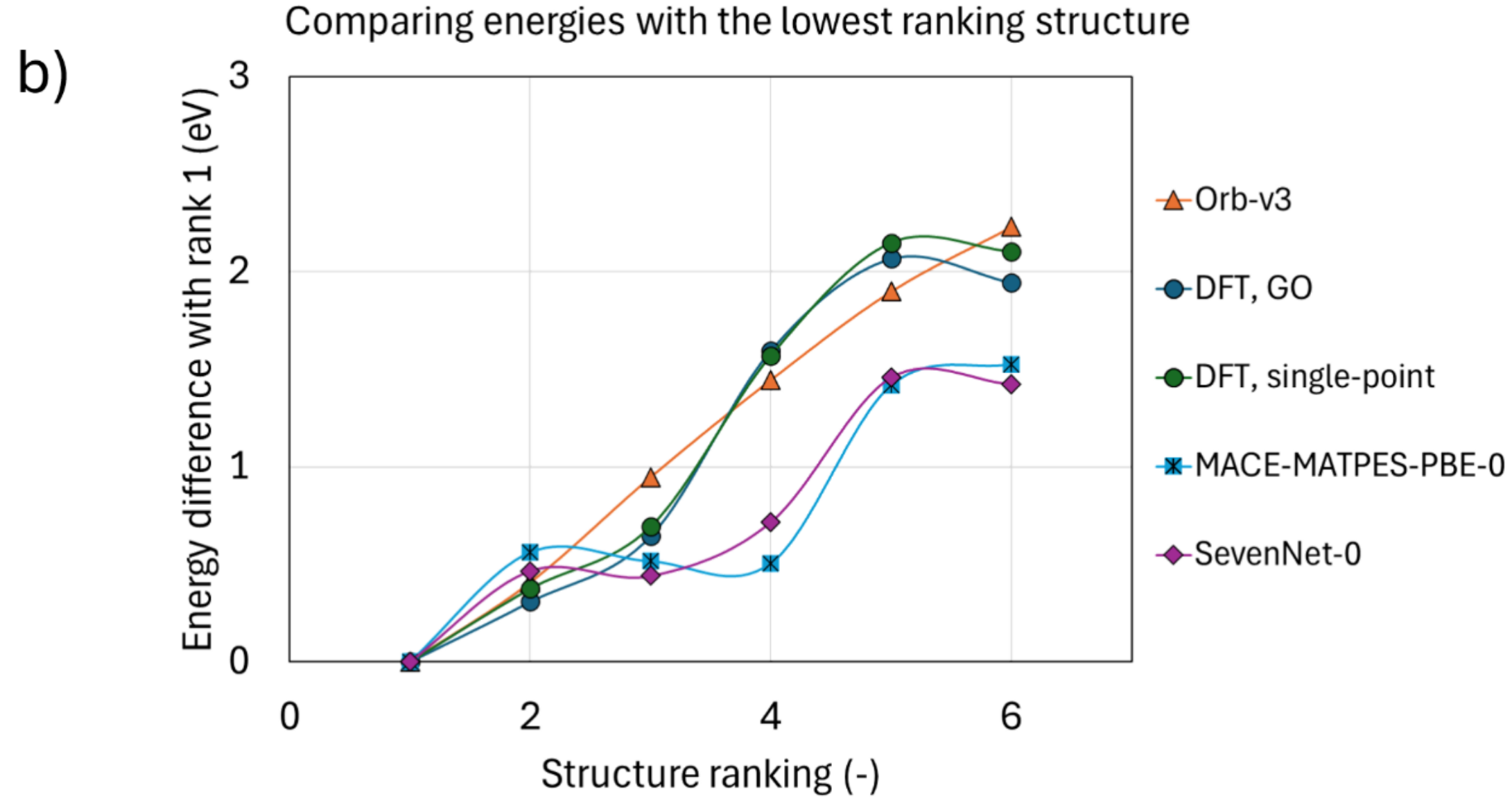

**Fig. 6: (a)** Comparison of Orb-v3 predictions with DFT single-point and fully geometry-optimized (GO) energies for six representative O interstitial configurations in Ti-23Nb-0.7Ta-2Zr. **(b)** Relative energies of all structures with respect to the lowest energy configuration predicted by Orb-v3, as obtained from the three uMLIPs and DFT. The ordering of structures follows the energy ranking determined by Orb-v3.

As mentioned in Section 3.1, SevenNet-0 deviates from MACE-MATPES-PBE-0 and Orb-v3 for H interstitials, predicting the octahedral interstitial sites to be energetically more favorable than the tetrahedral sites. To determine which relaxed final state is more consistent with DFT for the Ti-23Nb-0.7Ta-2Zr gum metal base alloy, we randomly selected six initial configurations with different local environments around the H interstitial. The final relaxed states obtained from MACE-MATPES-PBE-0 (stabilizing H in tetrahedral positions) and SevenNet-0 (stabilizing H in octahedral positions) were taken as the input structures for static (single-point) DFT calculations to directly compare their energetic stability.

As shown in **Fig. 7,** the structures with H occupying tetrahedral positions (from MACE-MATPES-PBE-0) exhibit significantly lower DFT energies than the corresponding structures predicted by SevenNet-0 (octahedral coordination), with energy differences ranging from 4.2 to 4.8 eV across all six configurations. These results indicate that SevenNet-0 converges to a qualitatively different local state that lies substantially higher in energy, highlighting its limitation in describing H behavior in this alloy.

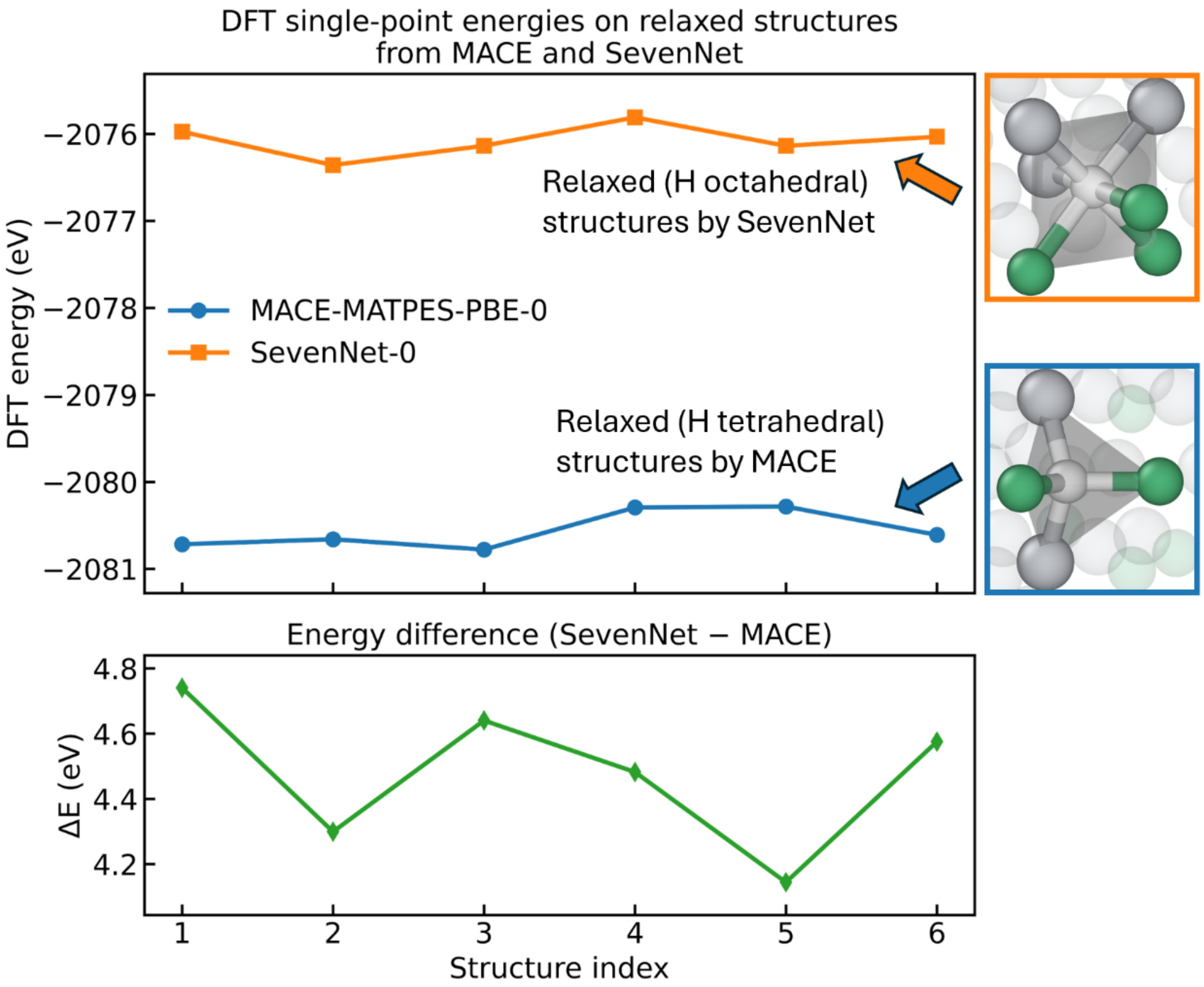

**Fig. 7:** Static (single-point) DFT energies evaluated for Ti-23Nb-0.7Ta-2Zr with H interstitial relaxed using two MLIPs: MACE-MATPES-PBE-0 (relaxing H into tetrahedral sites) and SevenNet-0 (relaxing H into octahedral sites). The lower panel shows the DFT energy differences between the two configurations. Across all structures, DFT energies are consistently lower for the MACE-relaxed states.

To further verify the stability of the respective configurations within DFT, we additionally performed full DFT geometry optimizations for three cases selected from the six configurations used in the static comparison. In each case, the DFT relaxation was initiated from both the MACE-relaxed (tetrahedral H) and SevenNet-relaxed (octahedral H) structures. During DFT optimization, the configurations initially stabilized by SevenNet in octahedral positions consistently transformed into tetrahedral positions, whereas the tetrahedral configurations obtained from MACE remained stable throughout the relaxation. The comparison of the energies from the static DFT on configurations as obtained from MACE-MATPES-PBE-0 and SevenNet-0 and after the geometry optimization are illustrated in **Fig. 8**. After geometry optimization, the DFT energies decrease further compared to the initial MACE structures. This additional energy lowering arises from minor adjustments in the positions of the alloy atoms, while the H atom remains in the tetrahedral site.

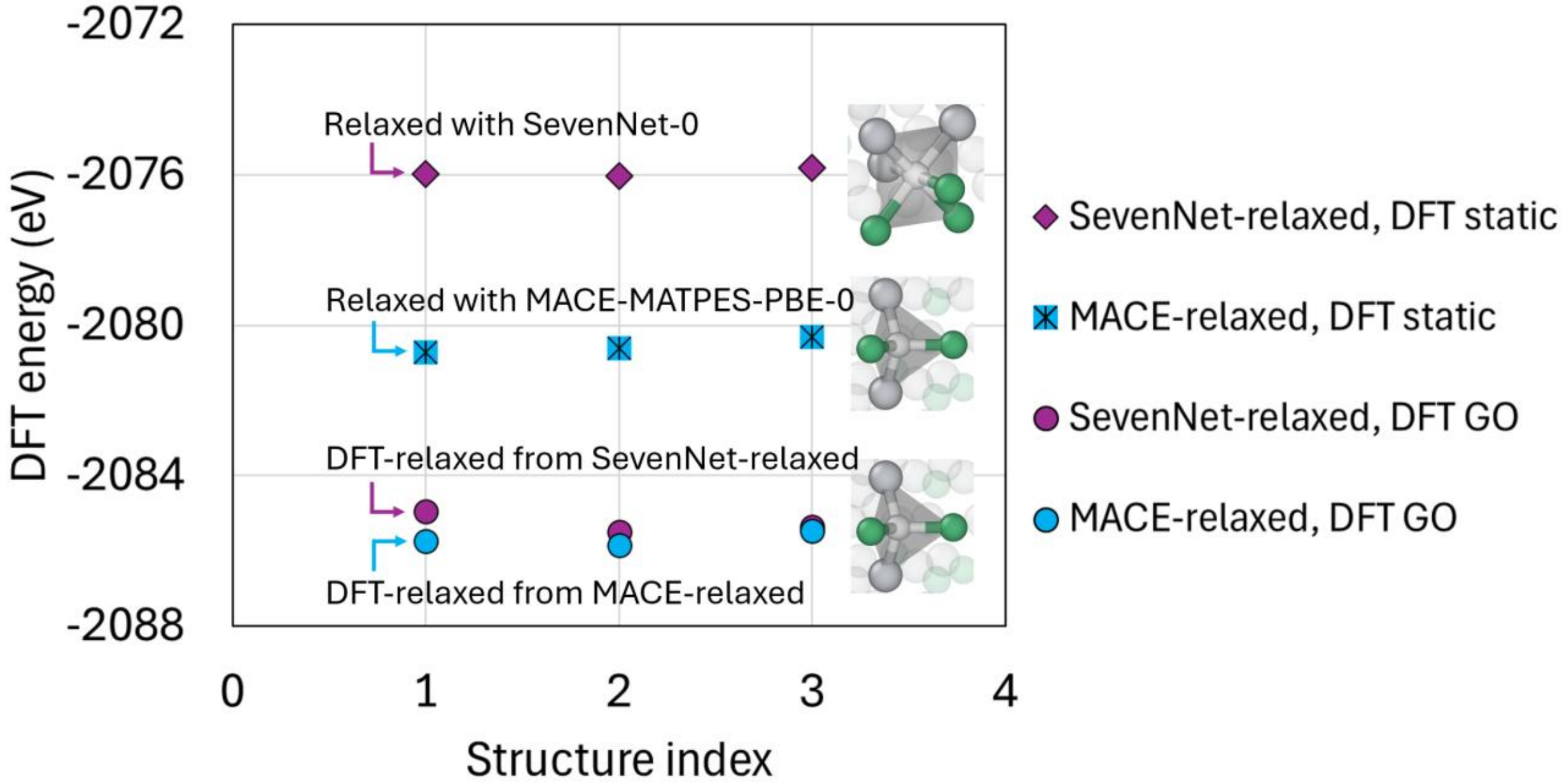

**Fig. 8:** Comparison of static and geometry-optimized (GO) DFT energies for 6 configurations of Ti-23Nb-0.7Ta-2Zr with H interstitial, where DFT relaxations were initiated from MACE- and SevenNet-relaxed structures, showing the transformation of SevenNet-stabilized octahedral H to tetrahedral sites and the stability of MACE-predicted tetrahedral H.

## 4. Discussion

### 4.1 Persistence of geometric site preferences

The preference of C, N, and O for octahedral sites and of H for tetrahedral sites with MACE-MATPES-PBE-0 and Orb-v3 is in agreement with the typical behavior in bcc alloys. C, N, and O consistently relax into octahedral sites, while H remains most stable in tetrahedral positions. The persistence of these preferences across thousands of chemical environments in our SQSs indicates that the geometrical preferences between octahedral and tetrahedral voids are not affected by local chemical environment differences. In other words, even though interstitial sites in the Ti-23Nb-0.7Ta-2Zr alloy differ in their chemical surroundings, the geometric character of the bcc lattice still governs the preferred coordination environment. These observations are consistent with prior studies on other multicomponent alloys. For instance, in TiNbZr MEA, Wu et al., 2025[16] reported preferential occupation of tetrahedral site for H. In HEA of HfNbTiVZr, Casillas-Trujillo et al, 2020[41] reported preferential occupation of octahedral sites for C.

SevenNet-0, however, illustrated a clear exception for H. While it correctly predicts octahedral minima for C, N, and O, it stabilizes H in octahedral rather than tetrahedral coordination. This behavior diverges from both the known physics of H in bcc alloys and the predictions of the other two uMLIPs, suggesting its possible limitation in describing correct H behavior in the studied gum metal base alloy. To assess which relaxed configuration is energetically consistent with DFT, we performed static DFT calculations on MLIP-relaxed structures obtained with MACE-MATPES-PBE-0 and SevenNet-0 (see **Fig. 7**). These results confirm a strong energetic preference for structures relaxed with MACE. Furthermore, full DFT relaxations (**Fig. 8**) consistently identify the tetrahedral configuration as the energetically favored site for H in the Ti-23Nb-0.7Ta-2Zr gum metal base alloy.

### 4.2 Chemical origins of Ti-driven stabilization and Nb-driven destabilization

Across all four interstitial elements (C, N, O, H), the dominant chemical trends revealed by the uMLIPs are the **(i)** strong energetic preference for Ti-rich local environments and the **(ii)** consistent penalty associated with Nb in close distance. These behaviors can be rationalized from electronic interactions between the host elements and the interstitials: Ti donates electronic charge to light interstitials and forms strong bonding interactions with them. This lowers the local energy and makes Ti-coordinated sites systematically more favorable. Generally, Zr behaves in a chemically similar way as Ti: it also forms strong bonds with these interstitials. However, having only 2 at.% Zr in the present gum-metal base alloy,

Zr-rich neighborhoods are statistically rare, and its stabilizing effect appears as statistically insignificant.

On the other hand, Nb shows the opposite behavior. Its more occupied d-states interact less favorably with the electronic states of the interstitials, resulting in weaker bonding. This leads to an energetic penalty when C, N, O, H are surrounded by Nb atoms. The destabilization is clearly visible through the distance descriptor: interstitial configurations become progressively less favorable as Nb is closer to the interstitial, and more favorable as this distance increases. Ta displays broadly similar electronic tendencies as Nb, but its effect is negligible here due to its very low concentration.

These chemical trends (Ti-driven stabilization and Nb-driven destabilization) are consistent with previous first-principles and machine-learning studies of M/HEA[16,19,42,43], which likewise report that light interstitials preferentially associate with Ti or Zr while avoiding Nb. The present work extends those observations to thousands of distinct local configurations in a realistic gum metal base alloy composition.

## 4.3 Reliability of uMLIPs

The DFT validation confirms that uMLIPs can capture well the energetics of interstitial configurations in a chemically complex alloy, even without system-specific training or fine-tuning. Their exceptional computational speed enables capabilities that are otherwise typically inaccessible to DFT, including:

- evaluating thousands of interstitial configurations,
- performing calculations on relatively large supercells (250 atoms in our case),
- obtaining statistically converged chemical trends.

The speed advantage is substantial. On a high-end retail RTX 4090 GPU, a full geometry optimization typically took less than 1 minute using uMLIPs, whereas the corresponding DFT optimization (starting even from the uMLIP-relaxed geometry) required up to 24 hours. This represents an increase in computational speed of a factor more than 1400. The small atomic displacements that we observed during the subsequent DFT relaxations further demonstrate that uMLIPs can provide high-quality initial structures and can speed up convergence of DFT calculation itself. For these reasons, the uMLIPs make high-statistics studies possible also without access to HPC resources.

While the overall consistency among the three uMLIPs for chemical environments is encouraging, the H site preference discrepancy with SevenNet-0 highlights the importance of cross-model comparison. SevenNet-0 predicts an octahedral H minimum, in contrast to both MACE-MATPES-PBE-0, Orb-v3, and typical bcc structure behavior. This deviation does not undermine SevenNet-0 performance for heavier interstitials (C, N, O), but it demonstrates its possible error in correct description of H element. The comparison in **Fig. 7** highlights this difference: the DFT-calculated energy difference between MACE-relaxed and SevenNet-relaxed H configurations is not a simple energy offset between models. Both relaxed final states are evaluated using the same single method (DFT) via static single-point energies, so the results directly reflect the energetic quality of the MLIP-predicted minima. Ideally, independent MLIP relaxations starting from the same initial structure should converge to the same (or very similar) low-energy minimum of the underlying DFT landscape. The fact that SevenNet-0 consistently converges to substantially higher-energy final states than MACE indicates that, for H in the studied Ti-23Nb-0.7Ta-2Zr gum metal base alloy, SevenNet-0 identifies a different and energetically less favorable configurations. For these reasons, validating uMLIP predictions against multiple models or against targeted DFT benchmarks remains an important task.

## 5. Conclusions

This work demonstrates that uMLIPs can reliably resolve the energetics of light interstitials (C, N, O, H) in the Ti-23Nb-0.7Ta-2Zr gum metal base alloy at a level of configurational sampling that is challenging and computationally demanding to reach with DFT. By evaluating total of 6750 interstitial configurations within three 250-atom SQSs using uMLIPs (MACE-MATPES-PBE-0, Orb-v3, SevenNet-0), we reveal clear chemical preferences: Ti-rich local environments strongly stabilize all four interstitial elements, whereas proximity to Nb leads to energy increase. Zr and Ta exhibit no statistically significant influence in our study, reflecting their low concentrations in the alloy. Despite the substitutional disorder, standard site preferences of the bcc lattice persist, with C, N, and O relaxing into octahedral sites and H stabilizing in tetrahedral sites. SevenNet-0 represents the only exception, consistently predicting the octahedral minimum for H, indicating a possible limitation of this foundation model.

Benchmarking randomly selected configurations against DFT confirms that the uMLIPs reproduce the energetic ordering of chemically distinct interstitial environments reasonably well. Together, these results demonstrate that pretrained uMLIPs can be used as efficient screening tools for statistically converged mapping of interstitial energetics in the Ti-23Nb-0.7Ta-2Zr gum metal base alloy at near-DFT accuracy and several orders of magnitude lower computational cost. Additionally, the chemical trends identified here offer potential guidance for tailoring interstitial behavior in this gum metal base alloy.

# 6. Declarations

## 6.1 Funding

This work was supported by the Technology Agency of the Czech Republic [grant No. TN02000069/003] and the Grant Agency of the Czech Technical University in Prague [grant No. SGS24/121/OHK2/3T/12]

## 6.2 Acknowledgements

The authors have no additional acknowledgements to report beyond the funding listed in the previous section.

# 7. Data availability

To ensure reproducibility of the results presented in this study, all scripts used to create the SQSs and to perform calculations with uMLIPs are publicly available in the following GitHub repository: github.com/bracerino/Ti-23Nb-0.7Ta-2Zr-Interstitials

The tool used to perform calculations with uMLIPs, *uMLIP-Interactive*, is publicly accessible at: github.com/bracerino/uMLIP-Interactive. Video tutorials demonstrating how to run calculations directly through its interface can be found here: youtu.be/xh98fQqKXaI?si=6KW1MqYNtPYEwiUR. A tutorial on using the all-in-one generated Python script is available at: youtu.be/w6hmvzC2J-8?si=4y0EYACdx3mwyeLa

# 8. Author contributions

M.L. and J.D. conceived the original idea. M.L. wrote the initial draft, developed all codes, performed all calculations, and analyzed the results presented in the article. J.D. and P.V. supervised the project and contributed to the execution of the calculations, analysis of results, and improvement of the manuscript. V.M. improved the manuscript and assisted in results evaluation.

# 9. Competing interest

The authors declare no competing interests.